\documentstyle[12pt,bezier]{article}
\begin{document}
\def \inbar{\vrule height1.5ex width.4pt depth0pt}
\def \xC{\relax\hbox{\kern.25em$\inbar\kern-.3em{\rm C}$}}
\def \xR{\relax{\rm I\kern-.18em R}}
\newcommand{\xZ}{Z \hspace{-.08in}Z}
\newcommand{\xbe}{\begin{equation}}
\newcommand{\xee}{\end{equation}}
\newcommand{\xbea}{\begin{eqnarray}}
\newcommand{\xeea}{\end{eqnarray}}
\newcommand{\xnn}{\nonumber}
\newcommand{\xkt}{\rangle}
\newcommand{\xbr}{\langle}
\newcommand{\xlll}{\left( }
\newcommand{\xrrr}{\right)}
\newcommand{\xcun}{\mbox{\footnotesize${\cal N}$}}
\title{Eigenvalue Problem for Schr\"odinger Operators and
Time-Dependent Harmonic Oscillator}
\author{Ali Mostafazadeh\thanks{E-mail: alimos@phys.ualberta.ca}\\ \\
Theoretical Physics Institute, University of Alberta, \\
Edmonton, Alberta T6G 2J1, Canada,\\
and\\
Department of Mathematics,  College of Arts and Sciences,\\
Ko\c{c} University, Istinye, 80860 Istanbul, Turkey\thanks{Address after July 1, 1997}}
\date{June 1997}
\maketitle

\begin{abstract}
It  is shown that  the eigenvalue problem for the Hamiltonians 
of the standard form, $H=p^2/(2m)+V(x)$, is equivalent to the 
classical  dynamical equation for certain harmonic oscillators 
with time-dependent frequency.  This is another indication of 
the central role played by time-dependent harmonic oscillators 
in quantum mechanics. The utility  of the known results for
eigenvalue problem in the solution of the dynamical equations
of a class of  time-dependent harmonic  oscillators is also
pointed out.
\end{abstract}

\newpage

\baselineskip=24pt

Recently there has been a growing interest in the study 
of the quantum dynamics, i.e., solution of the Schr\"odinger 
equation
	\xbe
	H\psi=i\hbar\frac{d\psi}{dt}\;,
	\label{sch-eq}
	\xee
for time-dependent harmonic oscillator \cite{oscillator},
	\xbe
	H=\frac{p^2}{2M(t)}+\frac{1}{2}\,M(t)\omega(t)^2x^2\;,
	\label{osc}
	\xee
and its generalizations \cite{oscillator-general}. The basic 
idea used in these studies is the invariant method of Lewis 
and Riesenfeld \cite{le-ri}. Although the results of 
Refs.~\cite{oscillator,oscillator-general} have direct 
relevance for the construction of the squeezed states 
which have potential physical applications,  contrary to 
the claims made by the authors, they do not yield exact 
solution of the Schr\"odinger equation (\ref{sch-eq}). In 
reality, what is being done \cite{oscillator} is to show that 
the general solution of the Schr\"odinger equation can be 
expressed in terms of the solutions of the classical equation 
of motion. This is a second order differential equation with 
variable coefficients whose exact solution is not known. 
The purpose of this article is to show that an exact solution 
of the dynamical equation for  time-dependent harmonic
oscillator  includes as a special  case the solution for the
eigenvalue problem for arbitrary time-independent 
Hamiltonians of the standard form  $H=p^2/(2m)+V(x)$, i.e.,
	\xbe
	\left[ -\frac{\hbar^2}{2m}\,\frac{\partial^2}{\partial x^2}+ 
	V(x)\right]\psi_E=E\,\psi_E\;.
	\label{eg-va-eq}
	\xee

It is well-known that the eigenvalue problem (\ref{eg-va-eq}) 
can be viewed as a variational problem for the energy 
\cite{optimization}
	\xbea
	{\cal E}[\psi]&:=&\frac{\xbr \psi|H|\psi\xkt}{\xbr 
	\psi|\psi\xkt}\:=\:\frac{1}{(N[\psi])^2}
	\int dx~ \psi^*\left(-\frac{\hbar^2}{2m}\,\psi''+
	V\psi\right)\;,
	\label{energy}\\
	(N[\psi])^2&:=&	\int dx~ \psi^*\psi\;,~~~~~\psi''\:=
	\:\frac{d^2\psi}{dx^2}\;,\xnn
	\xeea
i.e., Eq.~(\ref{eg-va-eq}) can be easily shown to be 
equivalent to
	\xbe
	\frac{\delta {\cal E}[\psi]}{\delta\psi^*(x)}=0\;,
	\label{opt}
	\xee
where $\delta/\delta\psi^*(x)$ denotes a functional derivative.
Performing a simple integration by parts, one can formulate 
the corresponding variational  problem as that of the action
 functional $S=\int L ~dx$, where $L$  is the Lagrangian 
	\xbe
	L:=\frac{1}{2(N[\psi])^2}\left(\psi^{*'}\psi'+
	\frac{m V(x)}{\hbar^2}\,\psi^*\psi\right)\;,
	\label{L}
	\xee
and a prime means $d/dx$. Note that for the dynamical 
system described by $L$,  $\psi$ and $x$ play the role of 
the `position' and  `time', respectively. Unfortunately,  due 
to the presence of $N$ on the right-hand side of (\ref{L}), 
$L$ is not in the standard form.

In order to remedy this problem, let us introduce the 
normalized wave functions $\Psi:=\psi/N[\psi]$. Then, 
$L$ is equivalent to the Lagrangian
	\xbe
	L':=\frac{1}{2}\left[\Psi^{*'}\Psi'+
	\frac{m V(x)}{\hbar^2}\,\Psi^*\Psi\right]
	+\lambda (1-\int dx~\Psi^*\Psi)\;,
	\label{L'}
	\xee
where $\lambda$ is a Lagrange multiplier enforcing the 
normalization constraint.

Switching to real variables one can easily identify 
(\ref{L'}) as the Lagrangian for a `time'-dependent 
two-dimensional circular harmonic oscillator \cite{exact}
with coordinates $(q_1,q_2)$ and frequency $\omega(x)$ 
given by 
	\[\Psi=:q_1+iq_2\;,~~~\omega^2=
	-\frac{m V(x)}{\hbar^2}\;.\]
In fact, without loss of generality, one can assume that 
$\psi$ and consequently $\Psi$ are real.\footnote{This is 
because the Schr\"odinger operator appearing in 
Eq.~(\ref{eg-va-eq}) is real and linear. } In this case, one 
has
	\xbe
	L'=\frac{1}{2}\left[\Psi^{'2}-\omega(x)^2\Psi^2\right]+
	\lambda (1-\int dx~\Psi^2)\;,
	\label{L'-1}
	\xee
This is precisely the Lagrangian for a one-dimensional 
harmonic oscillator of unit mass and `time'-dependent 
frequency $\omega(x)$ whose position $\Psi$ is 
constrained to satisfy $\Phi:=1-\int\Psi^2 dx=0$. The 
constraint $\Phi=0$ is a rather unusual global constraint 
demonstrating the global character of the eigenvalue 
problem.

There is another way of relating the eigenvalue problem 
(\ref{eg-va-eq}) to time-dependent harmonic oscillators 
which separates the local and global features of the 
eigenvalue problem. In order to outline this approach, let 
us define the following one-parameter family of the
actions $S_E[\Psi]=\int L_E(\Psi,\Psi',x) dx$ with
	\xbe
	L_E:=\frac{1}{2}\left[\Psi^{*'}\Psi'-\Omega_E^2(x)
	\Psi^*\Psi\right]\;,~~~~~E\in\xR\;,
	\label{L-E}
	\xee
and $\Omega_E^2:=m[E-V(x)]/\hbar^2$. One can easily 
show that the classical equation of motion corresponding 
to $S_E$ is identical with Eq.~(\ref{eg-va-eq}).  However, 
not for every $E$ are the solutions square integrable. The 
requirement of the square integrability of the classical 
solution $\Psi_E$ is equivalent to the condition
	\xbe
	N[\Psi_E]=1\;.
	\label{condi}
	\xee
Hence, the eigenvalues $E$ corresponds to those 
solutions, i.e., classical trajectories $\Psi_E=\Psi_E(x)$, 
which satisfy (\ref{condi}).

Again, one can see that $L_E$ is the Lagrangian 
for a time-dependent two-dimensional circular harmonic 
oscillator. Choosing, $\Psi$ real, one obtains a 
one-dimensional harmonic oscillator with time-dependent 
frequency $\Omega_E$, i.e.,
	\xbe
	L_E=\frac{1}{2}\left[\Psi^{'2}-\Omega_E^2(x)
	\Psi^2\right]\,,~~~~~
	\Omega_E^2:=\frac{m[E-V(x)]}{\hbar^2}\;.
	\label{L-E-1}
	\xee
Therefore, if one could solve the classical dynamics of 
such oscillators, one would be able to reduce the 
eigenvalue equation (\ref{eg-va-eq}) to an algebraic 
equation, namely Eq.~(\ref{condi}).

This observation clearly demonstrates the central role
 time-dependent  harmonic oscillators play in spectral 
analysis of one-dimensional Schr\"odinger operators. 
It also indicates the degree of difficulty of the exact 
solution of the classical and in view of 
Refs.~\cite{oscillator} the quantum dynamics of 
time-dependent harmonic oscillators.

Another interesting observation is to use the 
knowledge about the known solutions
of the eigenvalue problem (\ref{eg-va-eq}) to construct 
solutions of the classical dynamics of certain 
time-dependent oscillators. In order to do this, one 
needs to select a potential $V$ for which some or all 
of the eigenfunctions $\Psi_E$ are known.  These would 
directly yield certain  exact solutions of the classical 
equation of motion for (\ref{L-E-1}). For example, let 
$V=\omega_0^2x^2/2$ for some real constant 
$\omega_0$, then $E= (n+1/2) \hbar\omega_0$ and 
$\Psi_E$ are the known normalized eigenfunctions of  
the one-dimensional time-independent harmonic 
oscillator. In view of the preceding discussion, $\Psi_E$ 
are also exact solutions for the classical dynamical 
equations for `time'-dependent harmonic oscillators 
with frequency
	\xbe
	\Omega_n(x)=\sqrt{\frac{m\omega_0}{\hbar}(
	n+\frac{1}{2}-\frac{\omega_0 x^2}{2\hbar})}
	\;,~~~~n=0,1,2,\cdots\:,
	\label{omega}
	\xee
and unit mass. Note that the frequency $\Omega_n$ 
become imaginary for the classically forbidden region.

Similarly, one can use the results for other exactly 
solvable eigenvalue problems \cite{exact} to obtain 
exact solutions for the classical dynamical equations 
of the corresponding time-dependent harmonic 
oscillators. In view of the results of Ref.~\cite{oscillator} 
this would also yield the exact solution of the 
Schr\"odinger equation for these oscillators.

The same approach may also be applied for 
higher-dimensional eigenvalue problems. In this 
case, one can show the equivalence of the eigenvalue 
equation and  the classical field equation for  certain
non-interacting Euclidean scalar field theories with 
variable mass.

In conclusion, the time-dependent harmonic oscillators 
are shown to play a most prominent role in the spectral 
theory of  Schr\"odinger operators. The relationship 
between these two apparently distinct subjects can be 
used to yield exact solution for the classical 
dynamics of a class of time-dependent oscillators. The 
implications of this observation in the construction of 
squeezed states as outlined in Ref.~\cite{oscillator} 
awaits further investigation.

\end{document}